\documentclass[pra,aps,showpacs,twocolumn]{revtex4}
\usepackage{epsf}
\usepackage[dvips]{color}
\newcommand {\be}{\begin{equation}}
\newcommand {\ee}{\end{equation}}
\newcommand {\bea}{\begin{eqnarray}}
\newcommand {\eea}{\end{eqnarray}}
\newcommand {\mbf}{\mathbf}
\newcommand {\mrm}{\mathrm}

\begin{document}

\title{Limits of sympathetic cooling of fermions by zero temperature bosons due to particle losses}
\author{L.~D. Carr$^*$, T.~Bourdel, and Y. Castin\\}
\affiliation{Laboratoire Kastler Brossel, Ecole Normale
Sup\'erieure, 24 rue Lhomond, 75231 Paris CEDEX 05, France}
\date{\today}

\begin{abstract}

It has been suggested by Timmermans [Phys. Rev. Lett. {\bf 87},
240403  (2001)] that loss of fermions in a degenerate system
causes strong heating.  We address the fundamental limit imposed
by this loss on the temperature that may be obtained by
sympathetic cooling of fermions by bosons.  Both a quantum
Boltzmann equation and a quantum Boltzmann \emph{master} equation
are used to study the evolution of the occupation number
distribution. It is shown that, in the thermodynamic limit, the
Fermi gas cools to a minimal temperature $k_{{\rm
B}}T/\mu\propto(\gamma_{{\rm loss}}/\gamma_{{\rm coll}})^{0.44}$,
where $\gamma_{{\rm loss}}$ is a constant loss rate, $\gamma_{{\rm
coll}}$ is the bare fermion--boson collision rate not including the reduction
due to Fermi statistics, and $\mu\sim k_{{\rm
B}}T_{{\rm F}}$ is the chemical potential. It is demonstrated
that, beyond the thermodynamic limit, the discrete nature of the
momentum spectrum of the system can block cooling. The unusual
non-thermal nature of the number distribution is illustrated from
several points of view: the Fermi surface is distorted, and in the
region of zero momentum the number distribution can descend to
values significantly less than unity. Our model explicitly depends
on a constant evaporation rate, the value of which can strongly
affect the minimum temperature.
\end{abstract}


\maketitle

\section{Introduction}
\label{sec:intro}

Evaporative cooling has proven able to obtain degenerate Fermi
systems~\cite{jin3,truscott1,schreck1,granade2002,roati2002,ohara1,hadzibabic2002,hadzibabic2003}.
As polarized fermions cannot undergo s-wave collisions, it is
necessary to sympathetically cool with another species or spin
state.  In contrast to the case of bosons~\cite{ketterle1}, it has
been suggested that Fermi systems are highly sensitive to
loss~\cite{timmermans2}.  In this article, we will investigate
this question for the case of sympathetic cooling by an ideal zero
temperature Bose gas, in order to identify the fundamental limit
imposed by loss. Sympathetic cooling via a degenerate Bose gas is
indeed used in several present
experiments~\cite{truscott1,schreck1,roati2002,hadzibabic2003}.
The lowest experimentally obtained temperature to date is
$T/T_{\mrm{F}}\sim 0.05$~\cite{hadzibabic2003}, where
$T_{\mrm{F}}$ is the Fermi temperature.

Various theoretical groups have considered improved cooling
schemes~\cite{holland1}, including optimizing the evaporation
rate~\cite{wouters2002}, using different trapping
frequencies~\cite{onofrio1,presilla2003}, or using laser rather
than evaporative
cooling~\cite{santos1999b,idziaszek2001,idziaszek2003}. We will
restrict our investigation to a simple model of sympathetic
cooling of a gas in a box, in which it will be shown that a
minimum temperature arises naturally as a result of loss of
particles. The discrete nature of the energy spectrum of the
system can also be a limiting factor. It will be demonstrated,
from several points of view, that the occupation number
distribution is non-thermal in a non-trivial sense.
``Temperature'' and ``chemical potential'' will therefore be
defined based on the total number and energy of the fermions,
rather than on the equilibrium nature, or lack thereof, of their
distribution.

Our presentation is organized as follows.  In
Sec.~\ref{sec:physics}, we physically motivate our idealized model
in the context of present experiments. In Sec.~\ref{sec:model} a
quantum Boltzmann equation is derived under the assumption that
the density operator is gaussian in the fermionic field, which
permits the use of Wick's theorem, to study the evolution of the
mean number distribution and the temperature. In
Sec.~\ref{sec:qbe} this equation is investigated numerically in a
discrete system and both numerically and analytically in the
thermodynamic limit. In Sec.~\ref{sec:wick} the probability
distribution of the occupation numbers is examined without the
assumption of Wick's theorem: a quantum Boltzmann \emph{master}
equation~\cite{gardiner1997,jaksch1997} based on the secular
approximation~\cite{cohentannoudji1,cirac1996} is derived and
simulated via Monte Carlo methods. Finally, in
Sec.~\ref{sec:conclusion} we conclude.

\section{A Model for Sympathetic Cooling}
\label{sec:physics}

In order to understand the limits of sympathetic cooling imposed
by particle loss, we use an idealized theoretical model.  This
model contains the following assumptions. Firstly, the bosons form
a perfect reservoir: they are non-interacting, at zero
temperature, are not significantly reduced in number during the
total observation time, and when excited to non-zero momentum
states are removed from the system by evaporation. Secondly, the
Fermi gas is non-interacting: the fermions are all in the same
spin state so that there is no s-wave contribution to their
interactions, and, in the low temperature regime which will be
considered, the p-wave contribution is negligible.  Thirdly,
fermion--boson interactions are modeled by a contact potential
proportional to the s-wave scattering length. Fourthly, the system
is enclosed in a three-dimensional box with periodic boundary
conditions with sides of length $L_x,L_y,L_z$: the volume shall be
denoted as $L^3\equiv L_x L_y L_z$. Fifthly, the evolution of the
system is described by discrete evaporation time steps of duration
$t_e$, the end points of which represent a full removal of the
bosonic particles in the modes with non-zero momentum:
$1/t_e$ may be interpreted as the evaporation
rate.  Sixthly, a constant fermion loss rate is introduced.

In present experiments it is commonly assumed that the Fermi and
Bose gases are cooled down while remaining thermalized at a finite
and common temperature.  This requires a sufficiently large
collision rate between bosons and fermions as well as between
bosons, in comparison to the evaporation rate.  In our model, each
time an excited boson is created by interaction with the Fermi gas
it is removed sufficiently rapidly so that, even if an
\emph{interacting} Bose gas was considered, the bosons would not
have time to thermalize. Consequently, the Fermi and Bose gases
are never at the same temperature.  It is in fact advantageous to
maintain the Bose gas at a temperature much smaller than that of
the Fermi gas: it is only in this case that all collisions between
fermions and bosons are efficient, in the sense that they decrease
the energy of the fermions and therefore lead to cooling.  

In this respect, our model is not intended to closely represent
current experiments; rather, the goal of this study is to explore
the fundamental cooling limits due to loss in an ideal system.
However, we note that in an actual experiment the regime
considered in our model may be obtained if thermalization of the
bosons is avoided by a sufficiently strong evaporation rate. We
also note that cooling of fermions by a nearly pure condensate was
realized in a recent experiment~\cite{hadzibabic2003}.

\section{The Quantum Boltzmann Equation}
\label{sec:model}

Consider the Hamiltonian $H=H_0+V$, where $H_0$ is the kinetic
energy of the Fermi-Bose mixture and $V$ is the interaction
energy, such that
\bea
H_0\equiv\sum_{\mbf{k}}\frac{\hbar^2k^2}{2m_{\mrm{F}}}\hat{c}_{\mbf{k}}^{\dagger}\hat{c}_{\mbf{k}}
+\sum_{\mbf{q}}\frac{\hbar^2q^2}{2m_{\mrm{B}}}\hat{b}_{\mbf{q}}^{\dagger}\hat{b}_{\mbf{q}}
\, , \\
V\equiv\frac{g(t)}{L^3}
\sum_{\mbf{k}^{\prime},\mbf{k},\mbf{q}^{\prime}\neq\mbf{q}}
\hat{b}^{\dagger}_{\mbf{q}^{\prime}}\hat{b}_{\mbf{q}}
\hat{c}^{\dagger}_{\mbf{k}^{\prime}}\hat{c}_{\mbf{k}}
\delta_{\mbf{k}^{\prime}+\mbf{q}^{\prime},\mbf{k}+\mbf{q}} \, ,
\label{eqn:potential}\eea
\be g(t)\equiv g_0 \sin^2(t\pi/t_e)\, , \label{eqn:gfb} \ee where
$\hat{c}^{\dagger}$ ($\hat{b}^{\dagger}$) refers to the creation
operator, subject to the usual Fermi (Bose) commutator relations,
for a single fermion (boson) of momentum $\mbf{k}$ ($\mbf{q}$).
The coefficient $g_0\equiv 2\pi\hbar^2 a_s/m_r$, $a_s$ is the
s-wave scattering length for fermion--boson interactions, and
$m_r$ is the reduced mass. The choice of a continuous time dependence for
the interaction potential in Eq.~(~\ref{eqn:gfb})
avoids strong non-adiabatic effects.  In contrast, an abrupt
switching on and off of the interaction potential is a source of
heating, leading in particular to divergence of the mean kinetic
energy in the rate equations to follow.  A Feshbach resonance
could be used to achieve a time-dependent coupling of the form
suggested by Eq.~(\ref{eqn:g})~\cite{vogels1,Inguscio}.  Note that the case
$\mbf{q}=\mbf{q}'$ has been discarded in
Eq.~(\ref{eqn:potential}), as it gives a contribution $g
N_{\mrm{B}}N_{\mrm{F}}/L^3$ which has no effect on the dynamics.

In order to apply perturbation theory, it is required that $t_e$
be sufficiently small. For example, in the thermodynamic limit in
the Fermi degenerate regime, it is necessary that \be
t_e\gamma_{{\rm coll}}T/T_{\mrm{F}}\ll 1 \,
,\label{eqn:perturb}\ee where \be\gamma_{{\rm coll}}\equiv
\frac{3}{8}\sigma n_{\mrm{B}} v_{\mrm{F}}\,
,\,\label{eqn:gammadefine}\ee $\sigma \equiv 4\pi a_s^2$ is the
cross section for scattering between a boson and a fermion,
$n_{\mrm{B}}$ is the bosonic density, and $v_{\mrm{F}}$ is the
Fermi velocity, where $v_{\mrm{F}}\equiv \hbar(6\pi
n_{\mrm{F}})^{1/3}/m_{\mrm{F}}$ in this spin-polarized system,
with $n_{\mrm{F}}$ the density of fermions. The factor of
$T/T_{\mrm{F}}$ in Eq.~(\ref{eqn:perturb}) is due to Pauli
blocking. The factor of 3/8 in Eq.~(\ref{eqn:gammadefine}) has
been included to account for the reduced interaction strength due
to the choice of the temporal profile of $g(t)$. Note that
$\gamma_{{\rm coll}}$ is time-dependent for a non-zero loss rate,
since the density of fermions decreases. Consider the number
operator \be \zeta\equiv
\hat{c}_\mbf{k}^{\dagger}\hat{c}_\mbf{k}\, . \label{eqn:zeta}\ee
The mean value of $\zeta$ at time $(n+1)t_e$ may be written \be
\langle \zeta\rangle[(n+1)t_e]=
\mrm{Tr}[\tilde{\rho}(nt_e)\Lambda(nt_e)]\, ,\ee where \be
\Lambda(t)\equiv \tilde{U}^{\dagger}[(n+1)t_e\leftarrow
t]\,\zeta\, \tilde{U}[(n+1)t_e\leftarrow t]\,
,\label{eqn:lambda}\ee and $\tilde{\rho}(nt_e)$ and $\tilde{U}$
are the density operator and the evolution operator in the
interaction picture.  The operator $\Lambda$ satisfies the
equation of motion \be i\hbar
\frac{d}{dt}\Lambda=[\tilde{V}(t),\Lambda(t)]\, , \ee with the
``final'' condition $\Lambda[(n+1)t_e]=\zeta$.  This is equivalent
to the integral equation \be \Lambda(t)=\zeta+
\int_{t}^{(n+1)t_e}\frac{d\tau}{i\hbar}[\Lambda(\tau),\tilde{V}(\tau)]\,\label{eqn:lamint}
.\ee  A perturbative development of $\Lambda$ may then be obtained
by iteration of Eq.~(\ref{eqn:lamint}) in powers of $V$ as
follows: \bea \Lambda(nt_e)\simeq\zeta
+\int_{nt_e}^{(n+1)t_e} \frac{dt}{i\hbar}[\zeta,\tilde{V}(t)]\nonumber\\
+\int_{nt_e}^{(n+1)t_e}\frac{dt'}{i\hbar}\int_{t'}^{(n+1)t_e}
\frac{dt}{i\hbar} [[\zeta,\tilde{V}(t)],\tilde{V}(t')]\, .
\label{eqn:per} \eea One must then calculate the expectation value
of $\Lambda(nt_e)$ with respect to the state of the system after
the evaporation step at time $nt_e$, which is defined in the
interaction picture by \be
\tilde{\rho}(nt_e)=\tilde{\rho}_{\mrm{F}}(nt_e^-)\otimes|N_{\mrm{B}}:\mbf{q}=\mbf{0}\rangle\langle
N_{\mrm{B}}:\mbf{q}=\mbf{0}|\, , \label{eqn:dens1}\ee \be
\tilde{\rho}_{\mrm{F}}(nt_e^-)\equiv {\rm
Tr}_{\mrm{B}}[\tilde{\rho}(nt_e^-)]\, .\ee The fact that, after
evaporation, all remaining bosons are in $\mbf{q}=\mbf{0}$ was
used.  Note that the number of bosons has been assumed to remain
approximately constant during the evaporation process, in keeping
with the assumption of a perfect reservoir. Depletion of the
reservoir can only decrease the cooling efficiency.

The mean occupation number of the single particle state with
momentum $\mbf{k}$ is defined as \be
N_n(\mbf{k})=\mrm{Tr}[\hat{c}^{\dagger}_{\mbf{k}}\hat{c}_{\mbf{k}}\tilde{\rho}(nt_e)]\,
.\label{eqn:ndefine}\ee As shown in App.~\ref{app:qft}, the time
development of Eq.~(\ref{eqn:per}) plus the use of an approximate
Wick factorization leads to an approximate rate
equation for the occupation numbers which iteratively describes
the development of the system in evaporation steps of period
$t_e$:
\bea N_{n+1}(\mbf{k})&=&(1-\gamma_{{\rm loss}}\,t_e)N_n(\mbf{k}) \nonumber \\
&&+\sum_{\mbf{k}'}\mathcal{P}(\mbf{k}'\rightarrow \mbf{k})
N_n(\mbf{k}')[1-N_n(\mbf{k})] \nonumber \\
&&-\sum_{\mbf{k}'}\mathcal{P}(\mbf{k}\rightarrow \mbf{k}')
[1-N_n(\mbf{k}')]N_n(\mbf{k})\, ,\label{eqn:rate} \eea
where \bea &&\mathcal{P}(\mbf{k}\rightarrow\mbf{k}')\equiv
\frac{N_{\mrm{B}}}{\hbar^2
L^6}|g(\omega)|^2\, ,\label{eqn:pdef}\\
g(\omega) &\equiv&\int_0^{t_e}dt\,g(t) \exp(i\omega t)
=\frac{e^{\frac{i}{2}\omega t_e}
g_0\sin\left(\frac{t_e\omega}{2}\right)}
{\omega\left[1-\left(\frac{\omega t_e}{2\pi}\right)^2\right]}\,
,\label{eqn:g}\\
\hbar\omega&\equiv& \frac{\hbar^2 k'^2}{2m_{\mrm{F}}}
+\frac{\hbar^2 (\mbf{k}-\mbf{k}')^2}{2m_{\mrm{B}}}-\frac{\hbar^2
k^2}{2m_{\mrm{F}}}\, ,\label{eqn:omega}\eea and the integer $n$ is
the previous number of iterations. In the right-hand side of
Eq.~(\ref{eqn:rate}), the
second term represents the sum of probabilities of moving a
fermion into state $\mbf{k}$ while the third term is the sum of
probabilities of moving a fermion out of state $\mbf{k}$. A loss
term with a constant rate has also been introduced, under the
assumption $\gamma_{{\rm loss}}t_e\ll 1$. This describes, for
example, collisions with background gas particles present in
experiments. Equation~(\ref{eqn:rate}) is the central result of
this section and the basis of further study in this article. Its
validity is subject to the necessary condition that the
probability of departure from mode $\mbf{k}$ after an evaporation
cycle is \be
\sum_{\mbf{k}'\neq\mbf{k}}\mathcal{P}(\mbf{k}\rightarrow\mbf{k}')[1-N_n(\mbf{k}')]\ll
1\, ,\label{eqn:condition}\ee for all populated levels $\mbf{k}$.
A similar condition must hold for the probability of arrival.

When $t$ is expressed in units of $\gamma_{\mrm{coll}}^{-1}$, the
evolution of $N_n(\mbf{k})$ is ultimately governed by three
continuous dimensionless parameters,
$\gamma_{\mrm{loss}}/\gamma_{\mrm{coll}}$, $t_e
E_{\mrm{F}}(0)/\hbar$, and $\hbar\pi t_e/(m_{\mrm{F}}L^2)$, as
well as the ratio of masses $\alpha\equiv m_{\mrm{B}}/m{\mrm{F}}$,
which is fixed for a particular experiment.  As the goal of this
work is to study the ultimate limits of sympathetic cooling, the
minimum temperature in units of the chemical potential shall
be studied as a function of these parameters. Temperature and
chemical potential are defined with respect to the total number of
fermions and the total energy, as given by the standard sums over
$\mbf{k}$ of $N_n(\mbf{k})$ and
$(\hbar^2k^2/2m_{\mrm{F}})N_n(\mbf{k})$~\cite{landau3},
respectively, not with respect to the equilibrium nature, or lack
thereof, of the number distribution.

One may ask if there are higher order effects on
Eq.~(\ref{eqn:rate}) which are important. The third order term of
Eq.~(\ref{eqn:per}) has a vanishing contribution; however, the
fourth order term contains several physical effects.  Firstly, it
contains a correction to the Born approximation for the scattering
of a single fermion with a single boson; this correction is small
provided that $k_{\mrm{F}} a\ll 1$.  This is equivalent to the
weakly interacting regime, since $k_{\mrm{F}}\propto
n_{\mrm{F}}^{1/3}$. Secondly, a boson may interact with a fermion
and leave the condensate, undergo a subsequent interaction with a
second fermion, and enter a final momentum state $\mbf{q}''$.
There are then two possibilities: if $\mbf{q}''=\mbf{0}$, bosonic
stimulation occurs, and the contribution is proportional to
$[n_{\mrm{B}} |g(\omega)|^2]^2$; in contrast, the sum over all
final states $\mbf{q}'' \neq \mbf{0}$ has a contribution
proportional to $n_{\mrm{F}} n_{\mrm{B}}|g(\omega)|^4$.  The
former, which represents effective interactions between fermions
mediated by the bosonic reservoir, has been studied
elsewhere~\cite{albus1}, and is here neglected. The latter is
smaller than the former by a factor of $N_{\mrm{F}}/N_{\mrm{B}}\ll
1$.

\section{Study of the Quantum Boltzmann Equation}
\label{sec:qbe}

In the following, Eq.~(\ref{eqn:rate}) is studied with three
different approaches. In Sec.~\ref{ssec:finite} the evolution is
investigated in a discrete system via numerical integration.  In
Sec.~\ref{ssec:numerical} the thermodynamic limit is taken and a
second numerical study is made.  Finally, in
Sec.~\ref{ssec:analytical} the thermodynamic limit is treated
approximately and analytically under the assumption of an
equilibrium Fermi distribution.

\subsection{Time Evolution for a Finite System} \label{ssec:finite}

We first consider the evolution of a discrete, finite system which
evolves according to Eq.~(\ref{eqn:rate}).  There are two distinct
regimes of $t_e$. When the half-width $2\pi/t_e$ of the function $g(\omega)$ in
Eq.~(\ref{eqn:g}) is large compared to the spacing $\delta E$ between
the values of $\hbar\omega$ in (\ref{eqn:rate}),
the discrete nature of the spectrum of the system does not play a
significant role: this we term the \emph{continuous regime}.
A calculation of a typical $\delta E$ is presented in App.~\ref{appen:de}.  Using
the result of this calculation, the continuous regime 
may be defined explicitly by 
\be
\frac{\delta E \,\,t_e}{2\pi\hbar} \simeq
\frac{8}{3} \left(\frac{E_F t_e}{\hbar N_F}\right) \frac{(\alpha+1)^2}{4\alpha^2}
\ll 1\,  \,\label{eqn:thermo}\ee 
where $\alpha$ is the mass ratio $m_B/m_F$.
If, in addition, $N(\mbf{k})$
varies slowly with respect to the momentum level spacing, one may
furthermore take the thermodynamic limit, as shall be elucidated
in Sec.~\ref{ssec:numerical}.  For $\delta E \,\,t_e/2\pi\hbar\gg 1$,
$g(\omega)$ is no longer well-resolved
and the discreteness plays a strong role: this we term the
\emph{discrete regime}.

\begin{figure}[tp]
\begin{center}
\epsfysize=17cm \leavevmode \epsfbox{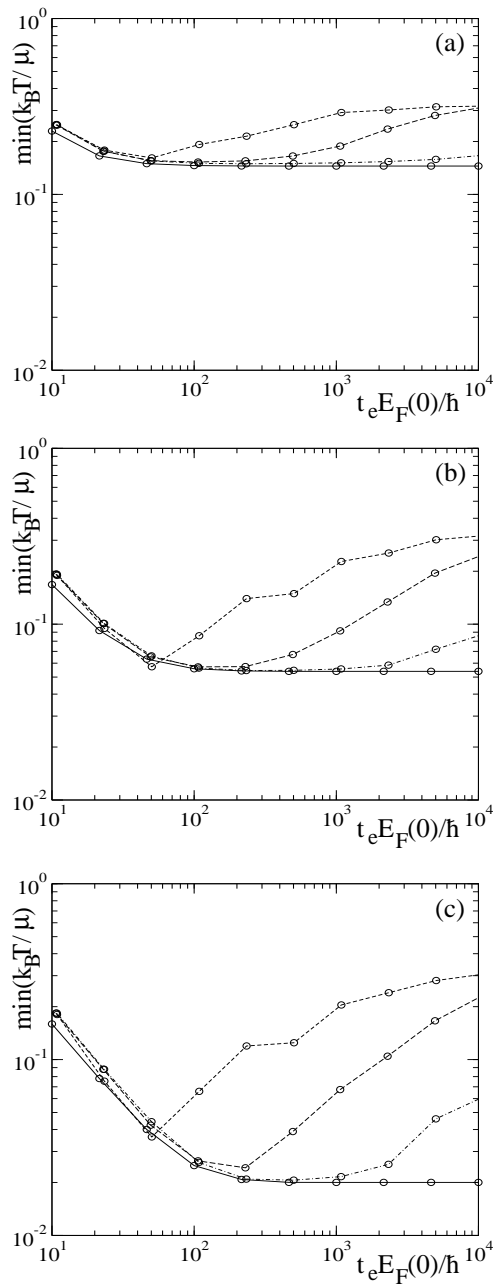}
\caption{\label{fig:1} Shown is the minimum value of
$k_{\mrm{B}}T/\mu$ as a function of the time step $t_e
E_{\mrm{F}}/\hbar$ for a $^7$Li -- $^6$Li mixture and
$\hbar\gamma_{\mrm{coll}}/E_{\mrm{F}}(0)=\frac{3}{8}\times
10^{-5}$. The ratio $\gamma_{\mrm{loss}}/\gamma_{\mrm{coll}}(0)$
is $8/3\times 10^{-2}$, $8/3\times 10^{-3}$, and $8/3\times
10^{-4}$ in panels (a), (b), and (c), respectively. The case of a
finite system, Eq.~(\ref{eqn:rate}), is shown for $10^2$ (dashed
line), $10^3$ (long dashed line), and $10^4$ (dot-dashed line)
fermions, and the thermodynamic limit, Eq.~(\ref{eqn:contrate}),
is shown as a solid line.  For the finite system, to the right
side of the plot, in the discrete regime, the cooling is blocked;
to the left, in the continuous regime, the minimum temperature is
dominated by the evaporation rate, according to
Eq.~(\ref{eqn:evap}), and rises as $t_e E_{\mrm{F}}/\hbar$
decreases. The data obtained in the thermodynamic limit are
independent of $t_e$ except where the evaporation-limited
temperature is larger than the loss-limited temperature.  Note that 
the actual data points are represented by open circles~\cite{fig1note}.}
\end{center}
\end{figure}

Figure~\ref{fig:1} shows a study of Eq.~(\ref{eqn:rate}) under
variation of the two central parameters
$\gamma_{\mrm{loss}}/\gamma_{\mrm{coll}}(0)$ and $t_e
E_{\mrm{F}}(0)/\hbar$ for the case of a $^{7}$Li -- $^{6}$Li
mixture, as in Refs.~\cite{schreck1,truscott1}.  Typical
experimental values of
$\gamma_{\mrm{loss}}/\gamma_{\mrm{coll}}(0)$ are $8/3\times
10^{-2}$ [Fig.~\ref{fig:1}(a)] to $8/3\times 10^{-3}$
[Fig.~\ref{fig:1}(b)].
The value $\gamma_{\mrm{loss}}/\gamma_{\mrm{coll}}(0)=8/3\times 10^{-4}$
[Fig.~\ref{fig:1}(c)], which could be reached by
using a Feshbach resonance~\cite{vogels1,Inguscio} to augment the
scattering length $a_s$ and thus the collision rate
$\gamma_{\mrm{coll}}\propto a_s^2$, was simulated as well.
Simulations of $N_{\mrm{F}}=10^2$ (dashed curve) and
$N_{\mrm{F}}=10^3$ (long dashed curve) fermions were performed in
a nearly cubic box with incommensurable sides; for computational
reasons, a cube was used for $N_{\mrm{F}}=10^4$ (dot-dashed
curve).  One clearly observes the continuous regime to the left
hand side of each plot, where the temperature is dominated by the
evaporation rate, as explained in Sec.~\ref{ssec:numerical}.  The
optimal temperature is achieved in the vicinity of $t_e
E_{\mrm{F}}/\hbar\sim 10^{2}$. Towards the right hand side of each
plot, the minimum temperature rises due to the blocking of cooling
by the discrete nature of the spectrum; this effect weakens for
larger numbers of atoms for a given value of $E_{\mrm{F}}t_e$ as
is apparent in Eq.~(\ref{eqn:thermo}). One may observe the
blocking in Eq.~(\ref{eqn:g}), where, for values of the energy
difference $\hbar\omega\gg\hbar/t_e$, $g(\omega)$ becomes small
and the interaction is effectively reduced. Finally, for values of
$t_e E_{\mrm{F}}/\hbar
> 10^{4}$ perturbation theory is no longer applicable, as
$\gamma_{\mrm{coll}}\hbar/E_{\mrm{F}}(0)=8/3\times 10^{-5}$ was
chosen, in keeping with typical experimental
conditions~\cite{typical}.

%
%

\begin{figure}[tp]
\begin{center}
\epsfysize=18cm \leavevmode \epsfbox{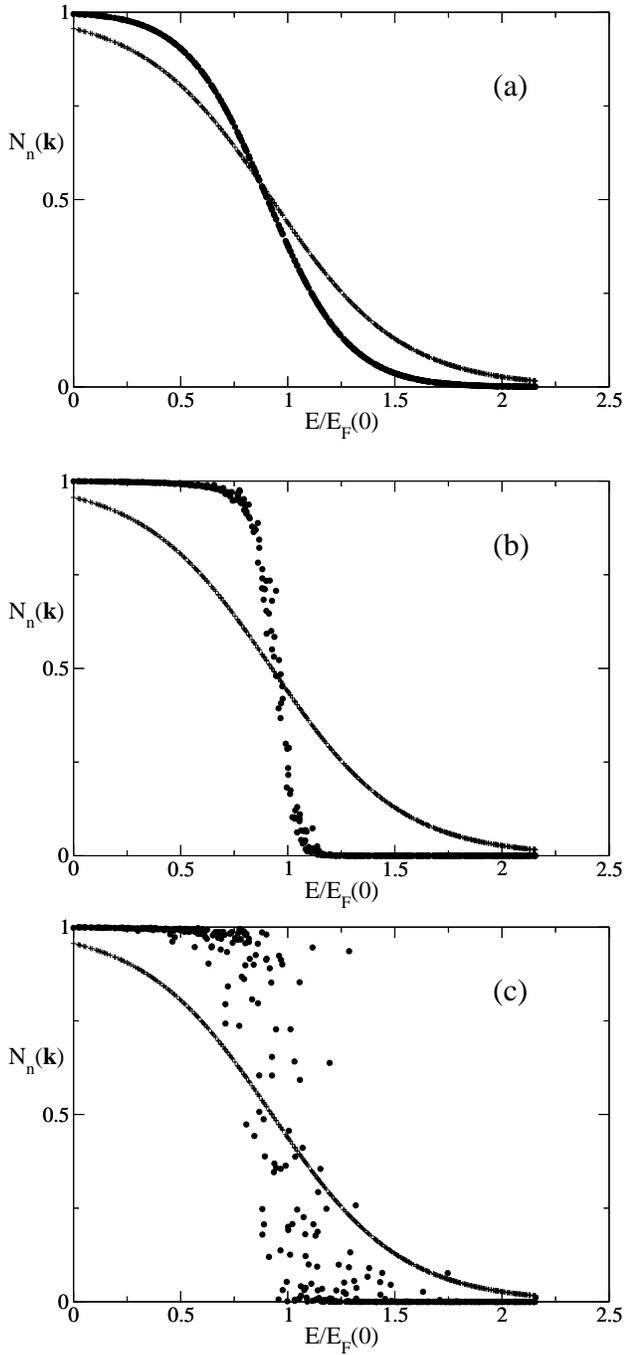}
\caption{\label{fig:2} Shown is the mean occupation number for
$N_{\mrm{F}}=1000$ $^6$Li atoms in contact with a $^7$Li reservoir
in a finite system. The initial state (plus symbols) is a thermal
distribution of $k_{\mrm{B}}T/\mu=0.324$;
$\hbar\gamma_{\mrm{coll}}/E_{\mrm{F}}(0)=\frac{3}{8}\times
10^{-5}$. After an evolution time (solid circles) 
such that $k_{\mrm{B}}T/\mu$
passes through a minimum according to Eq.~(\ref{eqn:rate}) with
the parameters 
$\gamma_{\mrm{loss}}/\gamma_{\mrm{coll}}(0)=8/3\times 10^{-3}$ and
(a) $t_e E_{\mrm{F}}/\hbar = 10^2$ (continuous regime) (b) $t_e
E_{\mrm{F}}/\hbar = 10^3$, and (c) $t_e E_{\mrm{F}}/\hbar = 10^4$
(discrete regime), it is clear that in the continuous regime the
distribution, though non-thermal, depends only on $|\mbf{k}|$. In
contrast, in the discrete regime the distribution is fully
$\mbf{k}$-dependent.}
\end{center}
\end{figure}

Figure~\ref{fig:2} shows the effect of $t_e$ on the occupation
number distribution in energy space.  In the continuous regime the
distribution clearly depends on energy alone, so that in the
thermodynamic limit $N(\mbf{k})=N(k)$.  The distribution is not,
however, an equilibrium one, as is more easily observed in the
thermodynamic limit (see below). In contrast, in the discrete
regime the distribution is fully $\mbf{k}$-dependent. Careful
observation of the figure shows regular holes in the energy
spectrum: this is a natural result of the quantization of the box.
We note that, in all regimes investigated, the essential feature
of a Fermi surface, though distorted, persists.

%
%

\subsection{Time Evolution in the Thermodynamic Limit}
\label{ssec:numerical}

%
\begin{figure}
\begin{center}
\epsfxsize=8cm \leavevmode \epsfbox{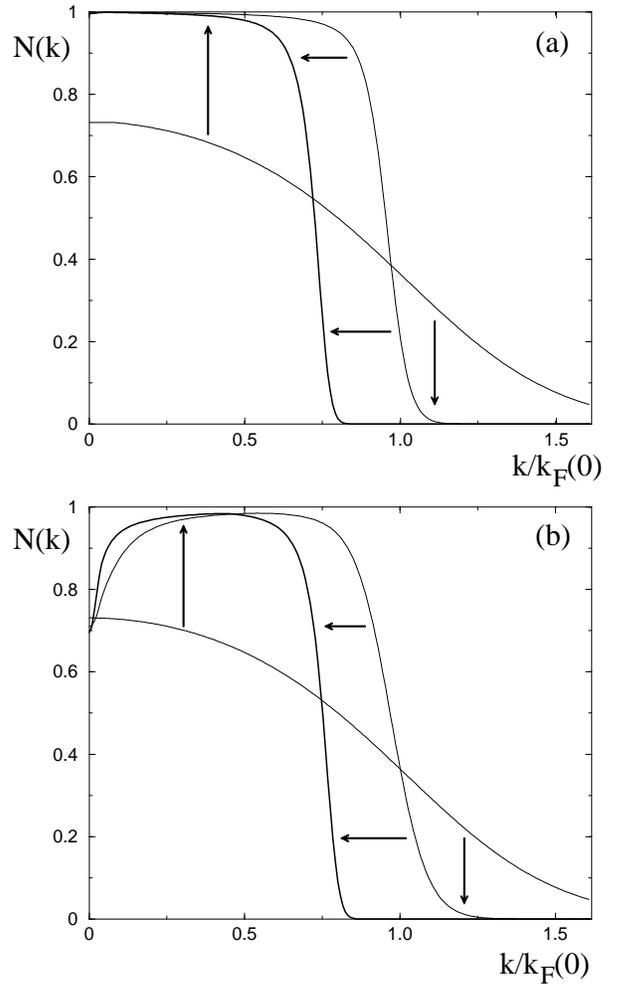}
\caption{\label{fig:3}  Time evolution of the number distribution
of a Fermi gas sympathetically cooled by a perfect Bose reservoir
for two experimentally important cases in the thermodynamic limit:
(a) $^7$Li -- $^6$Li and (b) $^{23}$Na -- $^6$Li. Shown are the
initial state of $(k_{\mrm{B}}T/\mu)(0)=0.7$ (thin line), the
state at which the temperature reaches a minimum (thicker line),
and a later stage where the gas has heated (thickest line).  The
hole in the distribution near $k=0$ is particularly evident in
panel (b), where the loss processes balance evaporative cooling,
as described by Eq.~(\ref{eqn:hole}). Here
$[\gamma_{\mrm{loss}}/\gamma_{\mrm{coll}}(0),(k_{\mrm{B}}T/\mu)_{\mrm{min}}]
= [8.27\times 10^{-3},0.081]$, and $[1.25\times 10^{-2},0.089]$
for panels (a) and (b), respectively, and
$\hbar\gamma_{\mrm{coll}}/E_{\mrm{F}}(0)=\frac{3}{8}\times
10^{-5}$.}
\end{center}
\end{figure}
%

In the thermodynamic limit, as defined explicitly in
Sec.~\ref{ssec:finite}, the sums in Eq.~(\ref{eqn:rate}) may be
approximated using the standard continuum limit \be
\sum_{\mbf{k}}\rightarrow L^3\int \frac{d\mbf{k}}{(2\pi)^3}\, .
\ee  As illustrated in Fig.~\ref{fig:2}, in this limit \be
N(\mbf{k})=N(k)\, . \ee The continuum iterative equation for the
time evolution may then be written as \bea
N_{n+1}(k)=(1-\gamma_{{\rm loss}}\,t_e)N_n(k)
+\frac{n_{\mrm{B}}}{\hbar^2}\int_0^{+\infty}dk'\,k'^2\nonumber\\
\times\left\{N_n(k)[1-N_n(k')]\int_{-1}^{+1}du |g(\omega)|^2 -
(k\leftrightarrow k')\right\}\, , \label{eqn:contrate} \eea where
$g(\omega)$ and $\omega$ are defined as in Eqs.~(\ref{eqn:g})
and~(\ref{eqn:omega}) with \be \mbf{k}\cdot\mbf{k}'=kk'u \, .\ee

Two effects limiting the choice of $t_e$ are implicit in
Eq.~(\ref{eqn:contrate}).  The first, which appears when $t_e$ is
small, is the width of the interaction function $g(\omega)$: as
$k_{\mrm{B}} T$ approaches the width $4\pi\hbar/t_e$ in the course
of cooling, the sharp decrease in $N_n(k)$ from unity to zero,
typical of a Fermi or Fermi-like distribution, is no longer
resolved, and the transfer of momentum from fermions to bosons
ceases to have any effect on the temperature.  This gives an
absolute minimum temperature of \be T_{\mrm{min}}^{\mrm{evap}}\sim
\frac{\hbar}{t_e}\, .\label{eqn:evap}\ee  The superscript ``evap''
refers to the fact that $1/t_e$ is the evaporation rate. The
second effect, which occurs in the limit of large $t_e$, is the
validity of perturbation theory, according to
Eq.~(\ref{eqn:perturb}). Since this limitation is imposed by our
use of perturbation theory it is not fundamental to sympathetic
cooling~\cite{technicalities}.
Therefore, in practice, within the context of our
model, one should choose an evaporation rate such that \be
\frac{T_{\mrm{F}}}{T}\ll t_e E_{\mrm{F}}/\hbar \ll
\frac{T_{\mrm{F}}}{T}\frac{E_{\mrm{F}}}{\gamma_{{\rm
coll}}\hbar}\, .\, \label{eqn:contlimit}\ee  We have verified that
the results are independent of values of $t_e$ which satisfy
Eq.~(\ref{eqn:contlimit}), as illustrated by the plateau in
Fig.~\ref{fig:1}.

It is convenient to begin with $N_0(k)$ in the form of a Fermi
distribution (though other initial distributions were studied,
with the same qualitative results). In the following simulations,
$k_{\mrm{B}}T/\mu = 0.7$ was taken as a starting condition. In
Fig.~\ref{fig:3} is shown the evolution of the occupation number
distribution resulting from Eq.~(\ref{eqn:contrate}) with a choice
of $t_e$ satisfying Eq.~(\ref{eqn:contlimit}) and
$\gamma_{\mrm{loss}}/\gamma_{\mrm{coll}}(0)= 8.27\times 10^{-3}$
and $1.25\times 10^{-2}$ for the experimentally relevant cases of
$^{6}$Li--$^{7}$Li and $^{6}$Li--$^{23}$Na, as shown in panels (a)
and (b), respectively.  Close inspection of the figure shows that
the distribution is non-equilibrium: the thermal tail is missing,
and, as can be seen by making a fit to a Fermi distribution (not
shown), the rise from $N_n(k)=0.5$ towards unity with decreasing
$k$ is less sharp than that of a Fermi distribution with the same
total energy and number of fermions. There is also a hole near
$k=0$, which is difficult to see in Fig.~\ref{fig:3}(a) but
appears strongly in Fig.~\ref{fig:3}(b). One may observe the
existence of this latter feature directly from
Eq.~(\ref{eqn:contrate}) as follows.

The evolution of $N_n(k=0)$ may be approximated by assuming that
$N_n(k)$ varies slowly near the origin, which allows one to
replace $N_n(k')$ with $N_n(0)$ in Eq.~(\ref{eqn:contrate}). The
condition that the value of $N_n(0)$ increase is then \bea
\frac{\gamma_{{\rm loss}}}{\gamma_{{\rm coll}}} &\lesssim&
\frac{4C}{3\pi}[1-N(0)]\left(\frac{\hbar}{\alpha E_F
t_e}\right)^{1/2}(1+\alpha^2)\nonumber\\
&&\times\left[\frac{1}{|1-\alpha|^{3/2}}-\frac{1}{(1+\alpha)^{3/2}}\right]\,
,\label{eqn:hole}\eea  where $\alpha\equiv
m_{\mrm{B}}/m_{\mrm{F}}$ and $C\equiv 1.860266\ldots$. It is
therefore directly apparent that for sufficient loss rates the
number distribution has a hole at $k=0$. Moreover, since a factor
of $[1-N_n(0)]$ enters into the condition, the number distribution
\emph{never} reaches unity at $k=0$ and the distribution is never
a Fermi distribution.  It was observed numerically that this
feature extends up to a finite $k$, the width of which varies
dynamically and increases as $m_{\mrm{B}}/m_{\mrm{F}}$ takes a
value largely different from unity. The deepest point occurs at
$k=0$, and the maximum in time of $N_n(0)$ is given by replacing
the less than or about equal to sign with an equal sign in
Eq.~(\ref{eqn:hole}).  The evolution of the hole for a
$^{23}$Na--$^6$Li mixture is particularly apparent, since
$\alpha\simeq 23/6$ is far from unity, which reduces the value of
the right hand side of Eq.~(\ref{eqn:hole}) for a fixed value of
$N_n(0)$, as illustrated in Fig.~\ref{fig:3}(b).

\begin{figure}
\begin{center}
\epsfxsize=8cm \leavevmode \epsfbox{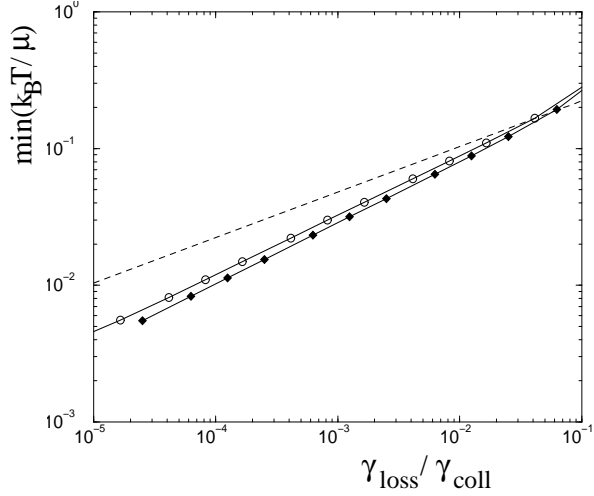}
\caption{\label{fig:4} The numerically determined maximum Fermi
degeneracy as a function of loss rate for $^7$Li -- $^6$Li (open
circles) and $^{23}$Na -- $^6$Li (black diamonds) mixtures in the
thermodynamic limit, according to the quantum Boltzmann equation.
One finds $(k_{\mrm{B}}T/\mu)_{\mrm{min}}
\propto(\gamma_{\mrm{loss}}/\gamma_{\mrm{coll}}(0))^{0.44}$.  The
dashed line shows the analytic prediction of
$(k_{\mrm{B}}T/\mu)_{\mrm{min}}
\propto(\gamma_{\mrm{loss}}/\gamma_{\mrm{coll}}(0))^{1/3}$ based
on the assumption of a Fermi distribution. The difference is due
to the non-thermal nature of the quasi-static mean occupation
number distribution, an example of which is illustrated in
Fig.~\ref{fig:3}. Here
$\hbar\gamma_{\mrm{coll}}/E_{\mrm{F}}(0)=\frac{3}{8}\times
10^{-5}.$}
\end{center}
\end{figure}

The effect of the choice of
$\gamma_{\mrm{loss}}/\gamma_{\mrm{coll}}$ on the minimal
temperature is shown in Fig.~\ref{fig:4}.  The power laws \bea \,
\left(\frac{k_{\mrm{B}}T}{\mu}\right)_{\mrm{min}}=
0.659\left(\frac{\gamma_{\mrm{loss}}}{\gamma_{\mrm{coll}}}\right)^{0.436}+5.6\times 10^{-4}\\
\label{eqn:powerlaw1}
\left(\frac{k_{\mrm{B}}T}{\mu}\right)_{\mrm{min}}=
0.621\left(\frac{\gamma_{\mrm{loss}}}{\gamma_{\mrm{coll}}}\right)^{0.445}+7.34\times
10^{-8}\label{eqn:powerlaw2}\eea for $^{7}$Na--$^6$Li  and
$^{23}$Na--$^6$Li, respectively, were found over the range
$8/3\times 10^{-5}\leq \gamma_{\mrm{loss}}/\gamma_{\mrm{coll}}\leq
8/3\times 10^{-2}$. Note that the constant offsets in the above
are negligible over the fit domain. Although the distributions
shown in Fig.~\ref{fig:3} are not Fermi distributions, the
step-like feature makes a temperature and chemical potential, as
defined by the total number of particles and the total energy, a
meaningful measure of the shape of $N_n(k)$.  Moreover, as
$N_{\mrm{F}}^{\mrm{tot}}$ and $E_{\mrm{F}}^{\mrm{tot}}$ are
weighted by $k^2$ and $k^4$ respectively, the hole near $k=0$ has
little effect on them.

%
%

%
\begin{figure}
\begin{center}
\epsfxsize=8cm \leavevmode \epsfbox{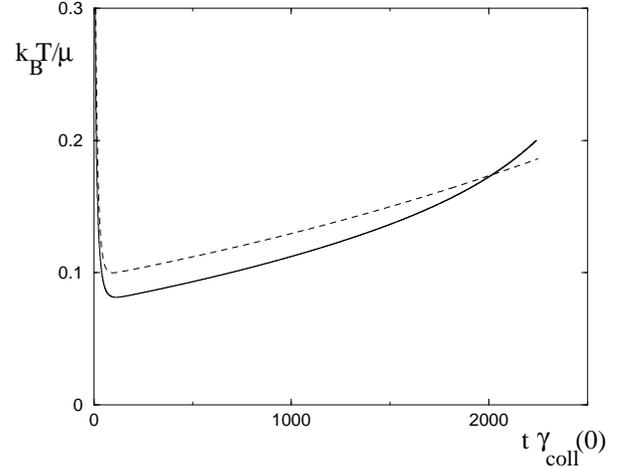}
\caption{\label{fig:5}  Time evolution of the degeneracy in the
thermodynamic limit as determined by numerical evolution of
Eq.~(\ref{eqn:contrate}) (solid line), and the analytical
prediction of Eq.~(\ref{eqn:tempevolve}) (dashed line) based on a
Fermi ansatz.  The parameters are
$\gamma_{\mrm{loss}}/\gamma_{\mrm{coll}}(0) = 8.27\times 10^{-3}$,
$(k_{\mrm{B}}T/\mu)(0)=0.7$, $t_e E_{\mrm{F}}(0)\hbar=10^3$, and
$\hbar\gamma_{\mrm{coll}}/E_{\mrm{F}}(0)=\frac{3}{8}\times
10^{-5}$.}
\end{center}
\end{figure}
%

Finally, Fig.~\ref{fig:5} illustrates the time evolution of
$k_{\mrm{B}}T/\mu$ for the parameters
$\gamma_{\mrm{loss}}/\gamma_{\mrm{coll}}(0) = 8.27\times 10^{-3}$
and $t_e E_{\mrm{F}}(0)\hbar=10^3$. The figure is divided into
three time regions which show three phases of the evolution:
cooling, the achievment of a minimal $k_{\mrm{B}}T/\mu$, and
heating. These three regions were observed in all simulations for
which $\gamma_{\mrm{loss}}/\gamma_{\mrm{coll}}\ll 1$, and were
independent of $t_e E_F/\hbar$ within the constraints of
Eq.~(\ref{eqn:contlimit}).

\subsection{Analytical Prediction of the Degeneracy for a Fermi Distribution Ansatz}
\label{ssec:analytical}

The following concerns the thermodynamic limit alone.  In the
limit in which \be \hbar/t_e\ll k_{\mrm{B}}T\, ,\,
\label{eqn:analyticlimit}\ee one may use the approximation \be
|g(\omega)|^2\sim \frac{3\pi}{4}g_0^2 t_e \delta(\omega)\,
,\label{eqn:delta}\ee where $\delta(\omega)$ is a Dirac delta
distribution. Substituting Eq.~(\ref{eqn:delta}) into
Eq.~(\ref{eqn:contrate}), and defining time continuously according
to \be \dot{N}(k,t)\equiv\frac{N_{n+1}-N_n}{t_e}\, , \,\ee the
iterative rate equation reduces to a first order partial
integro-differential equation with an integration over $k'$ alone:
\bea \dot{N}(k,t)=-\gamma_{{\rm loss}}N(k,t)+\frac{3}{8}\,\frac{n_{\mrm{B}} g_0^2 m_{\mrm{B}}}{2\pi\hbar^3}\nonumber\\
\times\left\{\int_k^{\left|\frac{1+\alpha}{1-\alpha}\right|k}dk'\,\frac{k'}{k}N(k',t)[1-N(k,t)]\right.\nonumber\\
-\left. \int_{\left|\frac{1-\alpha}{1+\alpha}\right|k}^k
dk'\,\frac{k'}{k}[1-N(k',t)]N(k,t)\right\}\, ,
\label{eqn:deltarate}\eea where $\alpha\equiv
m_{\mrm{B}}/m_{\mrm{F}}$. Here the functional form of the limits
of integration was determined by integration of the delta
distribution over the solid angle. Note that in the limit in which
$k\rightarrow 0$, $\dot{N}(k)= -\gamma_{{\rm loss}}N_n(k)$,
provided that $m_{\mrm{F}}\neq m_{\mrm{B}}$.  In the limit as
$m_{\mrm{B}}\rightarrow m_{\mrm{F}}\equiv m$, the limits of
integration of Eq.~(\ref{eqn:deltarate}) simplify to $[k,\infty]$
and $[0,k]$, respectively.  In this case the $1/k$ in the
integrand is not regulated and the expression diverges as
$k\rightarrow 0$, save in the case where $[1-N(k)]\rightarrow 0$.

An analytical model of the time evolution of the temperature can
be developed based on a Fermi distribution ansatz, namely, \be
N_a(k,t)\equiv\frac{1}{\exp\left\{\left[\frac{\hbar^2k^2}{2m}-\mu(t)\right]/k_{\mrm{B}}T(t)\right\}+1}\,
. \, \label{eqn:ansatz}\ee  For simplicity, the case $m\equiv
m_{\mrm{F}}=m_{\mrm{B}}$ is considered.  Two equations for the
unknowns $T(t)$ and $\mu(t)$ are obtained by multiplying
Eq.~(\ref{eqn:deltarate}) by $k^0$ and $k^2$ and integrating over
$\mbf{k}$: \bea
\frac{d}{dt}N_{\mrm{F}}^{\mrm{tot}}(t)&=&-\gamma_{\mrm{loss}}
N_{\mrm{F}}^{\mrm{tot}}\, , \,\label{eqn:nft}\\
\frac{d}{dt}E_{\mrm{F}}^{\mrm{tot}}(t)&=&\frac{L^3}{(2\pi)^3}\int
d^3k \frac{\hbar^2k^2}{2m} \dot{N}(k,t)\, ,\,\label{eqn:eft}
 \eea
where $N_{\mrm{F}}^{\mrm{tot}}$ and $E_{\mrm{F}}^{\mrm{tot}}$ are
the total number of fermions and total energy of the fermions,
respectively. In the degenerate regime where $k_{\mrm{B}}T/\mu\ll
1$, one may obtain an approximate evolution of $k_{\mrm{B}}T/\mu$
from Eqs.~(\ref{eqn:nft}) and~(\ref{eqn:eft}). One neglects terms
of order $\exp(-\mu/k_{\mrm{B}}T)$ in the right hand side of
Eq.~(\ref{eqn:eft}) and uses the low temperature expansions of
$N_{\mrm{F}}^{\mrm{tot}}$ and
$E_{\mrm{F}}^{\mrm{tot}}$~\cite{landau3}. Details are given in
App.~\ref{app:integral}. One finds
\be \frac{d}{dt}\left(\frac{k_{\mrm{B}}T}{\mu}\right)^2
=\frac{8}{5\pi^2}\gamma_{\mrm{loss}}
-\frac{12\zeta(3)}{\pi^2}\gamma_{\mrm{coll}}(t)\,\left(\frac{k_{\mrm{B}}T}{\mu}\right)^{3}\,
, \,\label{eqn:tempevolve}\ee
which clearly shows the separate contributions of heating due to
losses and cooling due to elastic collisions. The fact that the
cooling term is proportional to $T^3$ has a simple physical
interpretation.  Each collisional process occurs at a rate
$\gamma_{\mrm{coll}}k_{\mrm{B}}T/\mu$ due to Pauli blocking. It
involves a fraction $k_{\mrm{B}}T/\mu$ of the total number of
fermions.  The energy transferred to a boson per collisional event
is of order $k_{\mrm{B}}T$. Therefore the collisional term in
$dE/dt$ is proportional to $T^3$, from which follows
Eq.~(\ref{eqn:tempevolve}).

In the limit $\gamma_{\mrm{loss}}/\gamma_{\mrm{coll}}\ll 1$, which
is in fact the experimental case, one may distinguish three
different stages in the evolution of $k_{\mrm{B}}T/\mu$. In the
first stage, cooling dominates and it decreases according to the
power law \be
\left(\frac{k_{\mrm{B}}T}{\mu}\right)(t)\sim\frac{1}{6\zeta(3)\pi^{-2}\gamma_{\mrm{coll}}(0)\,t
+(\mu/k_{\mrm{B}}T)(0)}\, .\,\label{eqn:dstage1}\ee  Note that, in
the case where $\gamma_{\mrm{loss}}=0$, Eq.~(\ref{eqn:dstage1})
holds indefinitely. In the second stage, after a time
$t_{\mrm{min}}$, $k_{\mrm{B}}T/\mu$ arrives at a minima, given by
\bea
\left(\frac{k_{\mrm{B}}T}{\mu}\right)_{\mrm{min}}=\left[\frac{2}{15\zeta(3)}\frac{\gamma_{\mrm{loss}}}
{\gamma_{\mrm{coll}}(0)}\right]^{1/3}\, , \,\label{eqn:deltamin}
\\ t_{\mrm{min}}\simeq[\gamma_{\mrm{coll}}(0)]^{-2/3}\gamma_{\mrm{loss}}^{-1/3}\, . \,\eea
Note that $\gamma_{\mrm{loss}}t_{\mrm{min}}\ll 1$, so that a very
small fraction of the atoms have been lost when $k_{\mrm{B}}T/\mu$
achieves its minimum. In the third stage, heating manifests as an
adiabatic increase in $k_{\mrm{B}}T/\mu$, obtained by neglecting
$d(k_{\mrm{B}}T/\mu)^2/dt$ in Eq.~(\ref{eqn:deltarate}) and
thereby replacing $\gamma_{\mrm{coll}}(0)$ by
$\gamma_{\mrm{coll}}(t)=\gamma_{\mrm{coll}}(0)\exp(-\gamma_{\mrm{loss}}t/3)$
in Eq.~(\ref{eqn:deltamin}).  The evolution of $k_{\mrm{B}}T/\mu$
continues to increase adiabatically up till a characteristic time
given by \be
t_{\mrm{nonadiabatic}}\sim\frac{3}{\gamma_{\mrm{loss}}}
\ln\left(\frac{\gamma_{\mrm{coll}}}{\gamma_{\mrm{loss}}}\right)\,
.\ee Note that, at this evolution time, $k_{\mrm{B}}T/\mu$ is on
the order of unity, and the above analytical treatment ceases to
be applicable.

The number $\Delta N_{\mrm{B}}$ of bosons lost from the reservoir
during the first and second stages of cooling, {\it i.e.}, up till
the time at which the minimum $k_{\mrm{B}} T/\mu$ is achieved, may
be estimated simply from Eq.~(\ref{eqn:dstage1}).  One integrates
the rate of production of excited bosons $\propto
N_{\mrm{F}}(0)(k_{\mrm{B}}T/\mu)^2\gamma_{\mrm{coll}}(0)$ over
time, and finds the approximate relation \be \Delta
N_{\mrm{B}}\propto
N_{\mrm{F}}(0)\left(\frac{k_{\mrm{B}}T}{\mu}\right)(0).\ee  This
corresponds to the initial number of fermions active in the
cooling process.

In Fig.~\ref{fig:4} is shown the minimum temperature as a function
of loss rate.  The difference between the analytical prediction of
Eq.~(\ref{eqn:deltamin}) (dashed line) and the numerical results
of Eq.~(\ref{eqn:contrate}) (solid line) highlight the
non-equilibrium nature of the actual mean occupation number
distribution.  However, the qualitative features are the same for
both the analytical model and the numerical simulation.
Figure~\ref{fig:5} shows the evolution for the parameters
$\gamma_{\mrm{loss}}/\gamma_{\mrm{coll}}(0) = 8.27\times 10^{-3}$
and $(k_{\mrm{B}}T/\mu)(0)=0.7$.  The three stages of cooling,
achievement of a minimum temperature, and heating, are clearly
observable.  As $\gamma_{\mrm{loss}}/\gamma_{\mrm{coll}}(0)\ll 1$,
the time scale of the first stage is small compared to the third
stage.

\section{Beyond the Boltzmann approximation: the Master Equation}
\label{sec:wick}

The quantum Boltzmann equation is a closed equation obtained after
use of Wick's theorem to replace the mean value of occupation
numbers with the product of their mean values.  Such an
approximation applies when the probability distribution of the
density operator is nearly
Gaussian~\cite{gardiner1997,jaksch1997}, and neglects correlations
between modes.  In the following, the \emph{exact} probability
distribution of the occupation number shall be treated by deriving
a master equation for the fermion density operator in the limit of
weak coupling. Specifically, $g_0$ must be small enough so that
the probability to have more than one boson excited out of the
condensate during $t_e$ is negligible.  In the Fock basis, the
density operator is characterized both by `populations' and
`coherences', that is, by the diagonal and off-diagonal matrix
elements, respectively. Subsequently, the {\it secular
approximation} will be used to derive an equation for the
evolution of the distribution probability of the occupation
numbers.  This approximation applies in the regime in which the
evolution rate of the populations is much smaller than the Bohr
frequency of the coherences~\cite{cohentannoudji1}, which allows
one to derive closed equations for the populations.

\subsection{Derivation of the Quantum Boltzmann Master Equation} \label{ssec:master}

%
\begin{figure}
\begin{center}
\epsfxsize=8cm \leavevmode \epsfbox{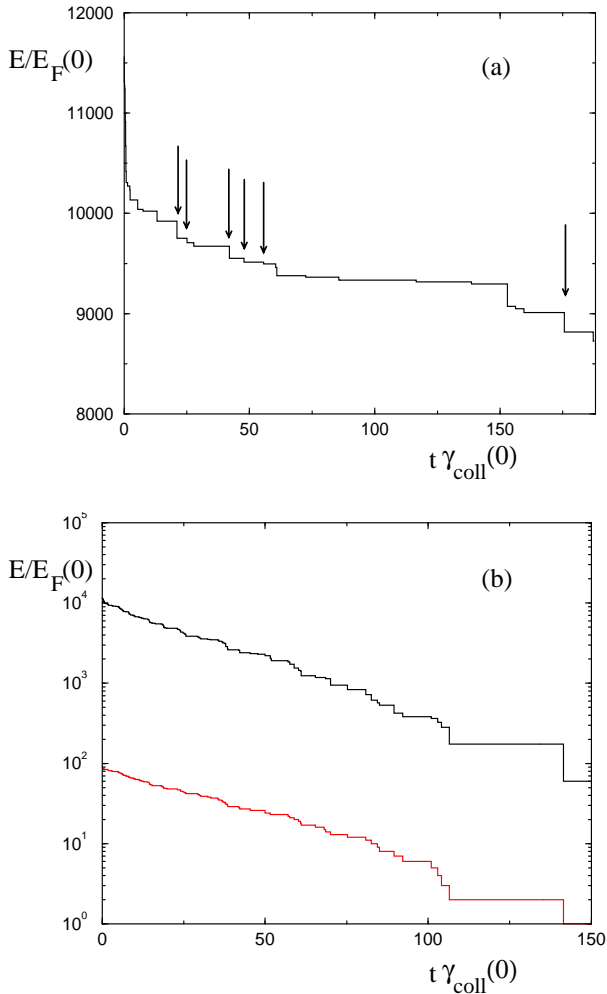}
\caption{\label{fig:6}  Evolution of the total energy of the
fermions for a finite system for one Monte-Carlo realization of
the \emph{exact} occupation number distribution under the secular
approximation (see text) according to Eq.~(\ref{eqn:qevolve}). (a)
$\gamma_{\mrm{loss}}/\gamma_{\mrm{coll}}(0) = 8/3\times 10^{-4}$:
the loss events (in this case 6) are marked by arrows.  Clearly
cooling dominates the decrease in the energy at early times while
loss plays a stronger role at later times.  (b)
$\gamma_{\mrm{loss}}/\gamma_{\mrm{coll}}(0) = 8/3\times 10^{-2}$:
The upper curve shows the evolution of the energy, and the lower
curve shows the evolution of the number of atoms.  At early times,
the slope of the energy is compounded of the effect of cooling and
a trivial decrease of the Fermi energy $\propto
\exp(-2\gamma_{\mrm{loss}}t/3)$ due to atom loss. In both (a) and
(b), $(k_{\mrm{B}}T/\mu)(0)=0.324$, $N_{\mrm{F}}(0)=100$, $t_e
E_{\mrm{F}}/\hbar=10^3$.}
\end{center}
\end{figure}
%

%
\begin{figure}
\begin{center}
\epsfxsize=8cm \leavevmode \epsfbox{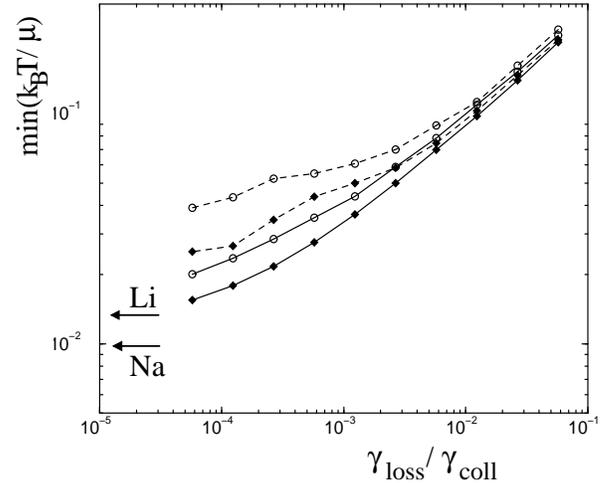}
\caption{\label{fig:7} The numerically determined maximum Fermi
degeneracy as a function of loss rate for $N=100$ (dashed line)
and $N=1000$ (solid line) $^6$Li fermions in contact with $^7$Li
(open circles) or $^{23}$Na (black diamonds), according to the
quantum \emph{master} Boltzmann equation. The value of $t_e
E_{\mrm{F}}/\hbar$ which gives the minimal temperature has been
chosen. Arrows indicate the long term cooling limits for
$\gamma_{\mrm{loss}}=0$ for $^7$Li--$^6$Li and $^{23}$Na--$^6$Li
mixes, respectively. The minimal temperature is much higher in
comparison to that predicted by the quantum Boltzmann equation in
the thermodynamic limit, as was shown in Fig.~\ref{fig:4} (note
scale change).}
\end{center}
\end{figure}
%

Just after a measurement of state of the bosons, but before
excited bosons have been removed from the system via evaporation,
the fermion density operator may be written \bea
\tilde{\rho}_{\mrm{F}}[(n+1)t_e]=\langle N_{\mrm{B}}:\mbf{0}|
\tilde{\rho}[(n+1)t_e]|N_{\mrm{B}}:\mbf{0}\rangle\nonumber \\
+\sum_{\mbf{q}\neq\mbf{0}}\langle N_{\mrm{B}}-1:\mbf{0};1:\mbf{q}|
\tilde{\rho}[(n+1)t_e]|N_{\mrm{B}}-1:\mbf{0};1:\mbf{q}\rangle\,
,\label{eqn:dens2}\eea where a sum over bosonic Fock states has
been taken. By choosing $g_0$ sufficiently small one may indeed
neglect the possibility of exciting more than one boson out of the
condensate.  The first term in Eq.~(\ref{eqn:dens2}) represents
zero excited bosons, and the second term a single excited boson of
momentum $\hbar\mbf{q}$.  The combined bosonic and fermionic
density operator $\tilde{\rho}[(n+1)t_e]$ is given by the action
on Eq.~(\ref{eqn:dens1}) of the evolution operator
$\tilde{U}(nt_e)$ from time $nt_e$ to time $(n+1)t_e$:
\bea \label{eqn:dens3}&\tilde{\rho}[(n+1)t_e]=\\
&\tilde{U}(nt_e)\left[\tilde{\rho}_{\mrm{F}}(nt_e^{-}) \otimes
|N_{\mrm{B}}:\mbf{q}=\mbf{0}\rangle \langle
N_{\mrm{B}}:\mbf{q}=\mbf{0}| \right]\tilde{U}^{\dagger}(nt_e)\,
.\nonumber\eea One then expands the evolution operator
$\tilde{U}(nt_e)$ to second order in the interaction potential $V$
using standard time-dependent perturbation theory, which results
in the second order expansion of Eq.~(\ref{eqn:dens3}), and thus
Eq.~(\ref{eqn:dens2}).

One obtains the following operators which act on the fermions
alone: \bea C_{\mbf{q}}(nt_e)&\equiv&
\int_{nt_e}^{(n+1)t_e}\frac{dt'}{i\hbar}\nonumber\\
&&\langle N_{\mrm{B}}-1:\mbf{0},1:\mbf{q}|\tilde{V}(t')
|N_{\mrm{B}}:\mbf{0}\rangle\, ,\label{eqn:c}\eea \bea
A(nt_e)&\equiv&\int_{nt_e}^{(n+1)t_e}\frac{dt'}{i\hbar}
\int_{nt_e}^{t'}\frac{dt''}{i\hbar}\nonumber\\
&&\langle N_{\mrm{B}}:\mbf{0}| \tilde{V}(t')\tilde{V}(t'')
|N_{\mrm{B}}:\mbf{0}\rangle\, .\label{eqn:a}\eea $C_{\mbf{q}}$ may
be written explicitly as \bea
C_{\mbf{q}}(nt_e)=\frac{\sqrt{N_{\mrm{B}}}}{i\hbar} \sum_{\mbf{k}}
g(\omega)e^{i\omega n t_e}
c^{\dagger}_{\mbf{k}-\mbf{q}}c_{\mbf{k}}\, ,\\
\omega\equiv \frac{\hbar
(\mbf{k}-\mbf{q})^2}{2m_{\mrm{F}}}+\frac{\hbar
q^2}{2m_{\mrm{B}}}-\frac{\hbar k^2}{2m_{\mrm{F}}}\,
.\label{eqn:omega2}\eea The explicit form of $A$ is more
complicated, but may be derived by substituting
Eq.~(\ref{eqn:potential}) into Eq.~(\ref{eqn:a}). $C_{\mbf{q}}$
originates from the second term of Eq.~(\ref{eqn:dens2}) in which
one boson is excited.  It originates from a first order
perturbative expansion of $\tilde{U}$, but appears in two factors
of Eq.~(\ref{eqn:dens3}) and therefore gives a contribution of
order $g_0^2$ in the evolution of the density operator. $A$
originates from the first term in Eq.~(43) with the evolution
operator expanded to second order in $g_0$. It contains, in
particular, an effective interaction between fermions mediated by
the bosons.  This results in the master equation for the fermionic
density operator \bea
\tilde{\rho}_{\mrm{F}}[(n+1)t_e]=\tilde{\rho}_{\mrm{F}}(nt_e)
+A(nt_e)\tilde{\rho}_{\mrm{F}}(nt_e)\nonumber\\
+\tilde{\rho}_{\mrm{F}}(nt_e)A^{\dagger}(nt_e)
+\sum_{\mbf{q}\neq\mbf{0}}C_{\mbf{q}}(nt_e)
\tilde{\rho}_{\mrm{F}}(nt_e)C^{\dagger}_{\mbf{q}}(nt_e)\,
.\label{eqn:pilote} \eea

We now proceed to apply the secular
approximation~\cite{cohentannoudji1}.  To illustrate the details,
the contribution of $C_{\mbf{q}}$ is described explicitly.
Evaluating the last term in Eq.~(\ref{eqn:pilote}), \bea
[C_{\mbf{q}} \tilde{\rho}_{\mrm{F}}C^{\dagger}_{\mbf{q}}](nt_e)
=\frac{N_{\mrm{B}}}{\hbar^2}\sum_{\mbf{k}_1,\mbf{k}_2} e^{i(\omega_1-\omega_2) nt_e}\nonumber\\
g(\omega_1)g^{*}(\omega_2)c^{\dagger}_{\mbf{k}_1-\mbf{q}}c_{\mbf{k}_1}
\tilde{\rho}_{\mrm{F}}(nt_e)
c^{\dagger}_{\mbf{k}_2}c_{\mbf{k}_2-\mbf{q}} \, ,\label{eqn:c2}
\eea where $\omega_{1,2}$ is defined as in Eq.~(\ref{eqn:omega2}),
with $\mbf{k}$ replaced by $\mbf{k}_{1,2}$. The typical evolution
rate $\gamma_{\mrm{evolve}}$ of the fermionic density operator is
proportional to $g_0^2$.  In contrast, the Bohr frequencies
$\omega_1-\omega_2$ do not depend on $g_0$. Therefore the
oscillating exponential in Eq.~(\ref{eqn:c2}) may be neglected for
sufficiently small $g_0$, as its effects averages to zero when
averaged during $1/\gamma_{\mrm{evolve}}$. An estimate for
$\gamma_{\mrm{evolve}}$ may be made based on the assumption of a
thermal distribution in the thermodynamic limit, as defined by
Eq.~(\ref{eqn:thermo}): \be \gamma_{\mrm{evolve}}=\frac{\delta
N_{\mrm{B}}}{t_e}
\simeq\gamma_{\mrm{coll}}N_{\mrm{F}}\left(\frac{T}{T_{\mrm{F}}}\right)^2\,
,\ee where $\delta N_{\mrm{B}}\ll 1$ is the mean number of excited
bosons during one cycle of duration $t_e$.  The minimal Bohr
frequencies are given by \be \mrm{min}(|\omega_1-\omega_2|)\simeq
\frac{1}{\hbar\rho(E_{\mrm{F}})}=\left(\frac{2\pi}{L}\right)^3
\frac{\hbar}{4\pi m k_{\mrm{F}}}\, ,\ee where $\rho(E_{\mrm{F}})$
is the density of states at the Fermi surface.  Therefore the
condition for validity of the secular approximation is \be
\frac{\hbar\gamma_{\mrm{coll}}}{E_{\mrm{F}}}N_{\mrm{F}}^2\left(\frac{T}{T_{\mrm{F}}}\right)^2\ll
1\, ,\ee for which, in Eq.~(\ref{eqn:c2}), one keeps only the
terms with $\omega_1=\omega_2$.  Assuming the box lengths squared
to be incommensurable, one finds \be
q_{\alpha}=0\:\:\:\mrm{or}\:\:\:k_{1\alpha}=k_{2\alpha}\, ,\ee for
each $\alpha\in\{x,y,z\}$.  The former case, which corresponds to
the excitation of a boson in the plane orthogonal to $\alpha$, we
neglect.  The existence of this solution is a consequence of the
separability of the degrees of motion along $x,y,z$ in the box.
One could consider an alternate model for the box in which this
separability is lifted to justify its being
neglected~\cite{santos1999b}.  There therefore remains the sole
condition $\mbf{k}_1=\mbf{k}_2$.

Having applied the secular approximation, if
$\tilde{\rho}_{\mrm{F}}$ is initially a statistical mixture of
Fock states, then it remains one for all times.  Defining the
occupation number probability distribution $Q_n(\{n_{\mbf{k}}\})$
by \be \tilde{\rho}_{\mrm{F}}(nt_e)=\sum_{\{n_{\mbf{k}}\}}
Q_n(\{n_{\mbf{k}}\}) |\{n_{\mbf{k}}\}\rangle\langle
\{n_{\mbf{k}}\}|\, , \,\ee where \be |\{n_{\mbf{k}}\}\rangle\equiv
\prod_{\mbf{k}}\left(\hat{c}_{\mbf{k}}^{\dagger}\right)^{n_{\mbf{k}}}
|\mrm{vac}\rangle\, ,\,\ee and each $n_{\mbf{k}}\in\{0,1\}$, one
obtains the equation of motion for $Q$:
\begin{widetext}
\bea &Q_{n+1}(\{n_{\mbf{k}}\})-Q_n(\{n_{\mbf{k}}\}) =
\sum_{\mbf{k_1}\neq\mbf{k_2}}\frac{N_{\mrm{B}}|g(\omega)|^2}
{\hbar^2 L^6}\left\{-n_{\mbf{k_1}}(1-n_{\mbf{k_2}})
Q_n(\{n_{\mbf{k}}\}) + n_{\mbf{k_2}}(1-n_{\mbf{k_1}})
Q_n(\{n_{\mbf{k}}
+\delta_{\mbf{k},\mbf{k_1}}-\delta_{\mbf{k},\mbf{k_2}}\})
\right\}\nonumber\\
&-\left(\sum_{\mbf{k_0}}\gamma_{\mrm{loss}} t_e
n_{\mbf{k_0}}\right) Q(\{n_{\mbf{k}}\})
+\left[\sum_{\mbf{k_0}}\gamma_{\mrm{loss}} t_e
(1-n_{\mbf{k_0}})\right]
Q(\{n_{\mbf{k}}+\delta_{\mbf{k},\mbf{k_0}}\})\,
,\,\label{eqn:qevolve}\eea
\end{widetext}
with \be \hbar\omega \equiv \frac{\hbar^2 k_2^2}{2m_{\mrm{F}}}
+\frac{\hbar^2
(\mbf{k_1}-\mbf{k_2})^2}{2m_{\mrm{B}}}-\frac{\hbar^2
k_1^2}{2m_{\mrm{F}}}\, .\,\ee  Here the loss term has been added
in by hand under the assumption that $\gamma_{\mrm{loss}}t_e
N_{\mrm{F}}\ll 1$.  The condition that the number of bosons
excited during $t_e$ be much smaller than unity may be written \be
\sum_{\mbf{k_1},\mbf{k_2}}\mathcal{P}(\mbf{k_1}\rightarrow\mbf{k_2})n_{\mbf{k_1}}(1-n_{\mbf{k_2}})\ll
1\, ,\ee for typical configurations $\{n_{\mbf{k}}\}$, where
$\mathcal{P}$ is defined as in Eq.~(\ref{eqn:pdef}). This is to be
contrasted with the much weaker condition of
Eq.~(\ref{eqn:condition}) obtained in the Quantum Boltzmann
equation, where the sum is over only one momentum.

\subsection{Monte Carlo Numerical Study} \label{ssec:monte}

The continuous time version of Eq.~(\ref{eqn:qevolve}), where
$Q_{n+1}-Q_n\equiv t_e \dot{Q}$, was studied numerically via Monte
Carlo simulation.  All data for $N_{\mrm{F}}(0)=10^2$ and
$N_{\mrm{F}}(0)=10^3$ in Fig.~\ref{fig:1} was re-evaluated, with a
mean over 100 realizations of $Q$ made for each data point. The
qualitative features of the evolution of the mean number
distribution were found to remain the same: its non-equilibrium
nature, as illustrated for the finite system in Fig.~\ref{fig:2}
and for the thermodynamic limit in Fig.~\ref{fig:3}; and the three
evolution stages of rapid cooling, achievement of a minimum, and
slow heating with a quasi-static distribution, as illustrated for
the thermodynamic limit in Fig.~\ref{fig:5}. Two examples of a
single Monte-Carlo realization are illustrated in
Fig.~\ref{fig:6}.  The minimum temperature for these simulations
is achieved at $\gamma_{\mrm{coll}}(0) \,t \sim 20$. Before this
time the evolution of the energy is dominated by cooling, rather
than losses: in Fig.~\ref{fig:6}(a) this is directly apparent, as
the loss of individual fermions is marked by arrows; in
Fig.~\ref{fig:6}(b), where the loss rate was higher, the slope is
seen to be different from that which results from the mere
evolution of the total number of fermions, the latter of which is
determined according to $\langle N_{\mrm{F}}^{\mrm{tot}}\rangle
=N_{\mrm{F}}^{\mrm{tot}}(0)\exp(-\gamma_{\mrm{loss}}t)$.  After
this time loss dominates, as is apparent in both panels: the
greater part of the changes in the energy occur at the same time
as loss of a fermion.

However, quantitatively the agreement between the results of the
quantum Boltzmann equation and the quantum Boltzmann master
equation depends both on the number of atoms and on the
evaporation rate. For 100 atoms, the minimum temperature predicted
by the master equation, as shown in Fig.~\ref{fig:7}, is as much
as 50\% higher as for the same parameters as in Fig.~\ref{fig:1}.
This deviation is strongest for small loss rates, where the
cooling efficiency is limited by the blocking mechanism due to the
discrete nature of the spectrum, as discussed in
Sec.~\ref{ssec:finite}. For 1000 atoms, the agreement is very good
for the loss rates of Figs.~\ref{fig:1}(a) and~\ref{fig:1}(b)
while for Fig.~\ref{fig:1}(c) there is a 15\% increase in the
temperature. The power laws \bea \,
\left(\frac{k_{\mrm{B}}T}{\mu}\right)_{\mrm{min}}=
0.861\left(\frac{\gamma_{\mrm{loss}}}
{\gamma_{\mrm{coll}}}\right)^{0.503}+1.44\times 10^{-2}\\
\label{eqn:powerlaw3}
\left(\frac{k_{\mrm{B}}T}{\mu}\right)_{\mrm{min}}=
0.855\left(\frac{\gamma_{\mrm{loss}}}
{\gamma_{\mrm{coll}}}\right)^{0.518}+0.993\times
10^{-2}\label{eqn:powerlaw4}\eea for $^{7}$Li--$^6$Li  and
$^{23}$Na--$^6$Li, respectively, were found over the full range of
data shown in Fig.~\ref{fig:7} for $N=1000$.  Comparing to
Eqs.~(\ref{eqn:powerlaw1}) and~(\ref{eqn:powerlaw2}), which were
obtained in the thermodynamic limit with the quantum Boltzmann
equation, one observes that the exponents of the power law are
similar, whereas the constant offset is substantially different.
Recalling that the data of Fig.~\ref{fig:7} resulted from an
optimization over $t_e$, this suggests that even in the absence of
loss there is an absolute minimal temperature of
$k_{\mrm{B}}T/\mu\sim 10^{-2}$.  This was validated for 1000 atoms
by Monte Carlo simulations with a zero loss rate, as indicated by
arrows drawn along the left hand $y$-axis of the figure: for
$^7$Li--$^6$Li, $\mrm{min}(k_{\mrm{B}}T/\mu)=1.34\times 10^{-2}$;
for $^{23}$Na--$^6$Li, $\mrm{min}(k_{\mrm{B}}T/\mu)=0.98\times
10^{-2}$.

Therefore, by taking the actual probability distribution for the
occupation numbers into account, {\it i.e.}, by not assuming
near-thermal equilibrium according to Wick's theorem, it is found
that the minimum temperature \emph{increases} when loss does not
dominate the cooling. This again highlights the non-thermal nature
of this system. In contrast, when $N_{\mrm{F}}$ is large and the
loss rate is in the experimental range of $10^{-2}\gamma_{\mathrm{coll}}$ 
to $10^{-3}\gamma_{\mathrm{coll}}$,
the secular approximation shows that Wick's theorem is a valid
assumption and one may simply use the quantum Boltzmann equation.

\section{Conclusion}
\label{sec:conclusion}

We have investigated the effect of heating caused by loss of atoms
on the minimum temperature that may be achieved in a
sympathetically cooled Fermi gas.  The model of sympathetic
cooling that we have used is cyclical and consists of a sequence
of time intervals during which fermions are coupled to a
zero-temperature ideal Bose gas via binary atomic interactions. At
the end of each time interval the excited bosons are removed from
the system by evaporation.  The length of the time interval is
short enough that the bosons do not come into thermal equilibrium
with the fermions; this is in contrast to present experiments in
which the bosons and fermions are always in equilibrium with one
another. The cooling is balanced by a constant loss rate of
fermions, which, for example, can be caused by collisions with
background gas in the experimental apparatus, and can lead to
heating, as shown in Ref.~\cite{timmermans2}. The combination of
cooling and heating are modeled at several levels of theoretical
approximation.

First, a quantum Boltzmann equation describing the evolution of
the mean occupation number distribution was developed under the
assumption that the fermion density operator is nearly Gaussian,
{\it i.e.}, that Wick's theorem may be applied. The overall
minimum temperature, which was found to be best obtained in the
thermodynamic limit, was observed to follow the power law
$(k_{\mrm{B}}T/\mu)_{\mrm{min}}\sim
0.65(\gamma_{\mrm{loss}}/\gamma_{\mrm{coll}})^{0.44}$,
 so that
$(k_{\mrm{B}}T/\mu)_{\mrm{min}}\lesssim 0.03$ for
$\gamma_{\mrm{loss}}/\gamma_{\mrm{coll}}\leq 10^{-3}$, where
$\mu\sim k_{\mrm{B}}T_{\mrm{F}}$ is the chemical potential of the
fermions, $\gamma_{\mrm{coll}}$ is the bare fermion--boson collision
rate not including the reduction due to Fermi statistics, 
and $\gamma_{\mrm{loss}}$ is the constant fermion loss rate.
This value of $\gamma_{\mrm{loss}}/\gamma_{\mrm{coll}}$ is easily
achievable in present experiments, in particular by using a
Feshbach resonance~\cite{Inguscio}. The number distribution was
observed to have a distorted Fermi surface and a hole near
$\mbf{k}=0$.

A second theoretical perspective was developed based on the
secular approximation to a master equation, without the use of
Wick's theorem. Monte Carlo simulations of the resulting quantum
Boltzmann \emph{master} equation showed that in the limit of
experimentally reasonable values of
$\gamma_{\mrm{loss}}/\gamma_{\mrm{coll}}$ and the number of
fermions $N_{\mrm{F}}$, the first theoretical approach is indeed
valid. However, for values of
$\gamma_{\mrm{loss}}/\gamma_{\mrm{coll}}$ tending towards zero,
the master equation shows a substantially higher temperature.  In
the most extreme case studied of 100 atoms and
$\gamma_{\mrm{loss}}/\gamma_{\mrm{coll}} = 3\times 10^{-4}$, this
increase is 50\%.

A possible extension to this work is to add a harmonic trap and/or
to include interactions between bosons.  The assumption of a
perfect Bose reservoir is reasonable when the speed of sound is
much smaller than the Fermi velocity~\cite{timmermans1998}, as is
the case for weakly interacting condensates.  Although a harmonic
trap may change the predicted minimal temperature, the qualitative
results of this study are expected to be correct even for a
non-uniform system such as is found in present experiments on
Fermi--Bose mixtures.

{\bf Acknowledgments:} We would like to thank Jean Dalibard for
proposing this project, Christophe Salomon for useful discussions,
and Murray Holland for a very useful remark. This work was
supported by NSF grant no. MPS-DRF 0104447. Laboratoire Kastler
Brossel is a research unit of l'\'Ecole normale sup\'erieure and
of l'Universit\'e Pierre et Marie Curie, associated with CNRS.  We
acknowledge financial support from R\'egion Ile de France.

\appendix
\section{Evolution of the occupation numbers}
\label{app:qft}

In the following, we will calculate an approximation to the
variation of the expectation value of the number operator \be
\zeta_{0}=\hat{c}^\dagger_{\mbf{k_0}}\hat{c}_{\mbf{k_0}} \ee from
time $n t_e$ to time $(n+1) t_e$, that is, during the time
interval of duration $t_e$ between two successive measurements of
the state of the bosons.

One begins with the approximate evolution of the number operator
obtained using second order perturbation theory for $\Lambda$,
Eq.~(\ref{eqn:per}). One takes the expectation of
Eq.~(\ref{eqn:per}) with respect to the density operator
Eq.~(\ref{eqn:dens1}). All the bosons at time $n t_e$ are in the
state with zero momentum. In order to take the expectation value
first with respect to the bosons, it is convenient to rewrite the
interaction potential $V$ as \be \tilde{V}(t) =
\sum_{\mbf{q}\neq\mbf{q}'} W_{\mbf{q}\mbf{q}'}(t)
\hat{b}_{\mbf{q}'}^{\dagger}\hat{b}_{\mbf{q}} \ee where the $W$'s
are purely fermionic operators: \be W_{\mbf{q}'\mbf{q}}(t) =
\frac{g_0(t)}{L^3} \sum_{\mbf{k}\neq\mbf{k}'}
e^{i\Omega_{\mbf{k}'\mbf{q}',\mbf{k}\mbf{q}}t}
\hat{c}^\dagger_{\mbf{k}'}\hat{c}_{\mbf{k}}
\delta_{\mbf{k}+\mbf{q},\mbf{k}'+\mbf{q}'}\, , \ee with \be
\Omega_{\mbf{k}'\mbf{q}',\mbf{k}\mbf{q}} \equiv \frac{\hbar
k^{'2}}{2m_{\mrm{F}}}+ \frac{\hbar q^{'2}}{2m_{\mrm{B}}}
-\frac{\hbar k^2}{2m_{\mrm{F}}}-\frac{\hbar q^2}{2m_{\mrm{B}}}.
\ee Expanding the commutators in (\ref{eqn:dens1}) and using the
fact that $\tilde{V}$ does not contain terms with
$\mbf{q}=\mbf{q}'$, one obtains the following matrix elements and
their expressions: \bea
\langle N_{\mrm{B}}:\mbf{0}| \tilde{V}(t) | N_{\mrm{B}}:\mbf{0} \rangle = 0 \\
\langle N_{\mrm{B}}:\mbf{0}| \tilde{V}(t) \ldots \tilde{V}(t') |
N_{\mrm{B}}:\mbf{0} \rangle \nonumber\\ = N_{\mrm{B}}
\sum_{\mbf{q}'\neq \mbf{0}} W_{\mbf{0},\mbf{q}'}(t)\ldots
W_{\mbf{q}',\mbf{0}}(t') \eea where $(\ldots)$ may contain an
operator acting on the fermions alone. These results may be
interpreted physically as follows. The action of $\tilde{V}$ on a
pure Bose-Einstein condensate creates a state with $N_{\mrm{B}}-1$
ground state bosons and a single excited boson with a
non-vanishing momentum $\mbf{q}'$, since the terms with
$\mbf{q}=\mbf{q}'$ have been excluded from the expression of
$\tilde{V}$, as apparent in Eq.~(\ref{eqn:potential}). The
resulting excited state of the bosons is orthogonal to the initial
state, so that the term of Eq.~(\ref{eqn:per}) linear in $V$ has a
vanishing expectation value. A second action of $\tilde{V}$ gives
a non-zero contribution to the expectation value only if the
excited boson is scattered back into the condensate.

There are also terms where the factors involving the interaction
potential appear in reverse chronological order, such as
$\tilde{V}(t') \tilde{V}(t)$. These terms are Hermitian conjugates
of the terms in chronological order so that the final result reads
\bea \langle\zeta_0\rangle[(n+1)t_e]-\langle\zeta_0\rangle(nt_e)=
-\frac{N_{\mrm{B}}}{\hbar^2} \sum_{\mbf{q}'\neq \mbf{0}}
\int_{nt_e}^{(n+1)t_e} dt' \nonumber \\ \int_{t'}^{(n+1)t_e} dt\,
\langle [\zeta_0, W_{\mbf{0},\mbf{q}'}(t)]
W_{\mbf{q}',\mbf{0}}(t') \rangle + \mbox{c.c.}\, ,
\label{eqn:commuhc} \eea where the expectation value in the right
hand side is taken with respect to the fermion density operator
$\tilde{\rho}_{\mrm{F}}(nt_e)$.

Finally, one evaluates the commutator in (\ref{eqn:commuhc}). The
identity \be [ \hat{c}^\dagger_{\mbf{k_0}}\hat{c}_{\mbf{k_0}},
\hat{c}^\dagger_{\mbf{k}'} \hat{c}_{\mbf{k}}] =
\delta_{\mbf{k_0},\mbf{k}'} \hat{c}^\dagger_{\mbf{k_0}}
\hat{c}_{\mbf{k}} -\delta_{\mbf{k_0},\mbf{k}}
\hat{c}^\dagger_{\mbf{k}'}\hat{c}_{\mbf{k_0}}\, ,\, \ee which is a
direct consequence of the fermionic anticommutation relations, is
used. Observing that the conservation of momentum imposes
$\mbf{k}'=\mbf{k}+\mbf{q}'$ in the expression for
$W_{\mbf{0},\mbf{q}'}$, one obtains \bea
[\zeta_0,W_{\mbf{0},\mbf{q}'}(t)] = \frac{g_0(t)}{L^3} \left(
\hat{c}_{\mbf{k_0}}^\dagger
\hat{c}_{\mbf{k_0}-\mbf{q}'}e^{i\Omega_{\mbf{k_0}\mbf{0},
\mbf{k_0}-\mbf{q}'\mbf{q}'}t}\right.\nonumber
\\\left. -\hat{c}_{\mbf{k_0}+\mbf{q}'}^\dagger \hat{c}_{\mbf{k_0}}
e^{i\Omega_{\mbf{k_0}+\mbf{q}'\mbf{0},\mbf{k_0}\mbf{q}'}t}
\right)\,.\, \eea Multiplying this expression by
$W_{\mbf{q}',\mbf{0}}$ gives fourth degree equation in the
fermionic creation/annihilation operators: to obtain a closed
equation for the occupation numbers one performs a crucial
factorization approximation based on the Wick contraction rule.
This constitutes the weak point of the present approach, which was
explored by a more systematic treatment in Sec.~V. As
the system is spatially homogeneous, the mean value of the product
of a creation operator and an annihilation operator of different
momentum states vanishes. One is left with \bea &\langle
\hat{c}_{\mbf{k_0}}^\dagger \hat{c}_{\mbf{k_0}-\mbf{q}'}
\hat{c}^\dagger_{\mbf{k}-\mbf{q}'}\hat{c}_{\mbf{k}}\rangle
\simeq \nonumber \\
&\delta_{\mbf{k},\mbf{k_0}} N_n(\mbf{k_0})[1-N_n(\mbf{k_0}-\mbf{q}')] \\
&\langle \hat{c}_{\mbf{k_0}+\mbf{q}'}^\dagger\hat{c}_{\mbf{k_0}}
\hat{c}^\dagger_{\mbf{k}-\mbf{q}'}\hat{c}_{\mbf{k}}\rangle \simeq
\nonumber \\
 &\delta_{\mbf{k},\mbf{k_0}+\mbf{q}'}[1-N_n(\mbf{k_0})]
N_n(\mbf{k_0}+\mbf{q}') \eea where the occupation numbers
$N_n(\mbf{k})$ are defined by Eq.~(\ref{eqn:ndefine}) and the fact
that $\mbf{q}'\neq \mbf{0}$ has been used. Observing that \bea
\int_{n t_e}^{(n+1)t_e} dt'\, \int_{t'}^{(n+1)t_e} dt\ g_0(t)
g_0(t') e^{i\omega(t'-t)} +\mbox{c.c.} \nonumber \\
=|g(\omega)|^2\, , \:\: \eea where $g(\omega)$ is defined in
(\ref{eqn:g}), one obtains the desired identity (\ref{eqn:rate}).

\section{Level spacing in the quantum Boltzmann equation}
\label{appen:de}
Consider a given single particle level 
of wavevector $\mathbf{k}$ in the box.
In the quantum Boltzmann equation (\ref{eqn:rate}), this level
is coupled to all the other levels $\mathbf{k}'$.  We wish
to estimate the mean distance between the values of the corresponding 
energy mismatches $\hbar\omega$ given by Eq.~(\ref{eqn:omega}). One may then define the density
of these values of $\hbar \omega$ by
\begin{equation}
\sigma(E) \equiv \sum_{\mathbf{k}'} \delta(\hbar\omega-E).
\end{equation}
Since $|g(\omega)|^2$ is centered in $\omega=0$, one can restrict the 
density of the $\hbar\omega$'s to $E=0$. Furthermore, we approximate
the discrete sum in $\sigma(0)$ by an integral:
\begin{equation}
\sigma(0) \simeq \left(\frac{L}{2\pi}\right)^3
\int d^3\mathbf{k}'\, \delta(\hbar\omega).
\end{equation}
This integral can be calculated exactly by using spherical coordinates
and integrating first on the polar angle, then on the modulus $k'$:
\begin{equation}
\sigma(0)= \left(\frac{L}{2\pi}\right)^3 \frac{m_B k}{2\hbar^2} \left[
1-\left(\frac{\alpha-1}{\alpha+1}\right)^2 \right],
\end{equation}
where $\alpha=m_B/m_F$. Taking the typical value $k\simeq k_F$,
using the relation $N_F=(L/2\pi)^3 4\pi k_F^3/3$ and setting $\delta E=1/\sigma(0)$,
one obtains Eq.~(\ref{eqn:thermo}).

\section{Evolution of the temperature}
\label{app:integral}

In the following, Eq.~(\ref{eqn:tempevolve}), which describes the
time evolution of $k_{\mrm{B}}T/\mu$ under the assumptions
\bea \hbar/t_e\ll k_{\mrm{B}}T\, , \label{eqn:bassume1}\\
m\equiv m_{\mrm{F}}=m_{\mrm{B}}\, ,\label{eqn:bassume2}\\
(k_{\mrm{B}}T/\mu)^2 \ll 1\, ,\label{eqn:bassume3}\\
N(k,t)\equiv N_a(k,t)\, , \label{eqn:bassume4} \eea is derived
from Eqs.~(\ref{eqn:nft}) and~(\ref{eqn:eft}). These four
assumptions correspond to the use of Fermi's Golden Rule, equal
masses of fermions and bosons, high degeneracy, and an equilibrium
Fermi distribution, respectively.
 Substituting Eq.~(\ref{eqn:deltarate}) into the right hand side of
Eq.~(\ref{eqn:eft}), \bea \dot{E}_{\mrm{F}}^{\mrm{tot}}(t)
=-\gamma_{\mrm{loss}}E_{\mrm{F}}^{\mrm{tot}}
+\Upsilon\int_{0}^{\infty}\frac{dx}{2}\nonumber \\
\times\left\{
x[1-\tilde{N}_a(x,t)]\int_{x}^{\infty}\frac{dy}{2}\, \tilde{N}_a(y,t)\right.\nonumber \\
\left. -x
\tilde{N}_a(x,t)\int_{0}^{x}\frac{dy}{2}\,[1-\tilde{N}_a(y,t)]\right\}
\, ,\,\label{eqn:b1} \eea where \be \Upsilon\equiv
\left(\frac{L}{2\pi}\right)^3 \left(\frac{\hbar^2}{2m}\right)
\left(\frac{3}{8}\,\frac{n_{\mrm{B}} g_0^2
m_{\mrm{B}}}{2\pi\hbar^3}\right)
\left(\frac{2mk_{\mrm{B}}T}{\hbar^2}\right)^3 4\pi\, ,\ee the
integration variables $x \equiv (\hbar^2 k^2)/(2mk_{\mrm{B}}T)$,
$y \equiv (\hbar^2 k'^2)/(2mk_{\mrm{B}}T)$, and
$\tilde{N}_a(x,t)\equiv N_a(k,t)$, $\tilde{N}_a(y,t)\equiv
N_a(k',t)$. The last term in Eq.~(\ref{eqn:b1}), \be I_2\equiv
-\frac{\Upsilon}{4} \int_{0}^{\infty}dx\,x
N_a(x,t)\int_{0}^{x}dy\,[1-N_a(y,t)]\, ,\ee may be integrated by
parts: \be I_2= -\frac{\Upsilon}{4} \int_{0}^{\infty}dx\,
[1-N_a(x,t)]\int_{x}^{\infty}dy\,y N_a(y,t)\, .\ee  This reduces
the integral in Eq.~(\ref{eqn:b1}) to \be I\equiv
\frac{\Upsilon}{4} \int_{0}^{\infty}dx \int_{x}^{\infty}dy\,
[1-N_a(x,t)] N_a(y,t) (x-y) \, .\label{eqn:b2} \ee  Making the
substitutions $\chi\equiv x-\mu/(k_{\mrm{B}}T)$, $\psi\equiv
y-\mu/(k_{\mrm{B}}T)$ and replacing $N_a$ according to
Eq.~(\ref{eqn:ansatz}), one obtains \be I = \frac{\Upsilon}{4}
\int_{-\frac{\mu}{k_{\mrm{B}}T}}^{\infty}d\chi
\int_{\chi}^{\infty}d\psi\,
\frac{\chi-\psi}{(e^{\psi}+1)(e^{-\chi}+1)} \, .\label{eqn:b3} \ee
The key to resolving Eq.~(\ref{eqn:b3}) is to take
$-\mu/(k_{\mrm{B}}T)\rightarrow -\infty$ in the lower limit of the
first integral.  Note that this is consistent with
assumption~(\ref{eqn:bassume3}).  In this case, \be
-\frac{1}{2}\frac{4}{\Upsilon}I= \zeta(3)=1.20206\cdots\,
,\label{eqn:b4} \ee where $\zeta$ is the Riemann zeta
function~\cite{abramowitz1}.

What is the error involved in this approximation?  Defining the
neglected portion of Eq.~(\ref{eqn:b3}) as \be J=
\int_{-\infty}^{-\frac{\mu}{k_{\mrm{B}}T}}d\chi
\int_{\chi}^{\infty}d\psi\,
\frac{\chi-\psi}{(e^{\psi}+1)(e^{-\chi}+1)} \,
,\label{eqn:berror1} \ee it is apparent that the second integral
is evaluated over $\psi\in[\chi\leq -\mu/(k_{\mrm{B}}T)\ll
-1,\infty]$.  The leading contribution of this integral therefore
gives \be \int_{\chi}^{\infty}d\psi\,
\frac{\chi-\psi}{(e^{\psi}+1)} \sim -\frac{1}{2}\chi^2\,
.\label{eqn:berror2} \ee  Substituting Eq.~(\ref{eqn:berror2})
into Eq.~(\ref{eqn:berror1}), the error is \be J\sim
-\frac{1}{2}\left[\left(\frac{\mu}{k_{\mrm{B}}T}\right)^2
+2\frac{\mu}{k_{\mrm{B}}T}+2\right]e^{-\frac{\mu}{k_{\mrm{B}}T}}\,
.\label{eqn:berror3} \ee Therefore \bea
\dot{E}_{\mrm{F}}^{\mrm{tot}}(t)
=-\gamma_{\mrm{loss}}E_{\mrm{F}}^{\mrm{tot}}
-\frac{1}{2}\Upsilon\zeta(3)\nonumber\\
+\mathcal{O}\left[\left(\mu/k_{\mrm{B}}T\right)^2
e^{-\frac{\mu}{k_{\mrm{B}}T}}\right]\, .\label{eqn:b5}\eea The
high degeneracy expansions \bea N_{\mrm{F}}^{\mrm{tot}}&=&
\left(\frac{L}{2\pi}\right)^3 4\pi \frac{1}{2}
\left(\frac{2m\mu}{\hbar^2}\right)^{\frac{3}{2}}\frac{2}{3}\nonumber \\
&\times&\left\{1 + \frac{\pi^2}{8}
\left(\frac{k_{\mrm{B}}T}{\mu}\right)^{2} +
\mathcal{O}\left[\left(\frac{k_{\mrm{B}}T}{\mu}\right)^{4}\right]
\right\}\, ,\,\label{eqn:b6}\\
E_{\mrm{F}}^{\mrm{tot}}&=& \left(\frac{L}{2\pi}\right)^3
\frac{\hbar^2}{2m}4\pi \frac{1}{2}
\left(\frac{2m\mu}{\hbar^2}\right)^{\frac{5}{2}}\frac{2}{5}\nonumber \\
&\times&\left\{1 + \frac{5\pi^2}{8}
\left(\frac{k_{\mrm{B}}T}{\mu}\right)^{2} +
\mathcal{O}\left[\left(\frac{k_{\mrm{B}}T}{\mu}\right)^{4}\right]
\right\}\, ,\,\label{eqn:b7}\eea may be easily developed from the
treatment of Ref.~\cite{landau3}.  Substituting Eq.~(\ref{eqn:b6})
into Eq.~(\ref{eqn:nft}) and Eq.~(\ref{eqn:b7})  into
Eq.~(\ref{eqn:b5}), one may solve Eq.~(\ref{eqn:nft}) for $\mu(t)$
and substitute the resulting expression into Eq.~(\ref{eqn:eft})
to obtain the final result, $(k_{\mrm{B}}T/\mu)(t)$, as shown in
Eq.~(\ref{eqn:tempevolve}).


\begin{thebibliography}{10}
\bibitem{fig1note}  Note that the
simulation was performed in a box of length ratios
$L_x:L_y:L_z=1.10345:1.0:0.9205$, excepting the case of $10^4$
atoms, for which the box was cubic.
\end{thebibliography}

\end{document}